\documentclass[12pt]{article}
\usepackage[active]{srcltx}

\usepackage{amsbsy}
\usepackage{amsfonts}
\usepackage{amssymb}
\usepackage{amsmath}
\usepackage{graphicx}

\def\IZ{\mathbb{Z}}

\def\be{\begin{eqnarray}}
\def\ee{\end{eqnarray}}
\def\Tr{{\rm Tr}}

\def\id{\protect{{1 \kern-.28em {\rm l}}}}

\makeatletter
\renewcommand\section{\@startsection {section}{1}{\z@}%
                                   {-3.5ex \@plus -1ex \@minus -.2ex}%
                                   {2.3ex \@plus.2ex}%
                                   {\normalfont\large\bfseries}}
\renewcommand\subsection{\@startsection{subsection}{2}{\z@}%
                                   {-3.25ex\@plus -1ex \@minus -.2ex}%
                                   {1.5ex \@plus .2ex}%
                                   {\normalfont\normalsize\bfseries}}
\makeatother

\begin{document}

${}$

\thispagestyle{empty} 
\vspace{1cm}

\begin{center}

{\Large\bf
Dark energy from \\
\vskip .2cm
a renormalization group flow
}

 \vspace{0.8cm} {I.~Mocioiu,$\footnote{irina@phys.psu.edu}$
  R.~Roiban $\footnote{radu@phys.psu.edu}$
  }\\
 \vskip  0.2cm
\small
{\em
    Department of Physics, The Pennsylvania State University,
University Park, PA 16802, USA
}

\normalsize
\end{center}

 \vskip 0.4cm

\begin{abstract}

We present evidence that a special class of gravitationally-coupled hidden sectors,
in which conformal invariance is dynamically broken in a controlled way, exhibit 
the properties of dark energy. 
Such quantum field theories may appear while embedding the Standard Model 
in a more fundamental high energy theory. 
At late times, an effective dark energy field behaves similarly to an exponentially 
small cosmological constant while at early times its energy density partly 
tracks that of matter. 

\end{abstract}

\newpage

\section{Introduction}

What is the structure of  our Universe and what governs its evolution?  Extensive 
cosmological observations have shown that the Universe is mostly cold, dark and 
accelerating (for a review see {\it e.g.} \cite{Frieman:2008sn} and references therein). 
A large fraction of its current (and possibly future) energy density may 
be modeled by an effective negative-pressure fluid  named "dark energy". Despite 
intense investigation, the fundamental structure and origin of this fluid 
is, however, not well understood.

A simple cosmological constant term in Einstein's equations provides perhaps the 
least complicated model of dark energy. The smallness of the observed cosmological 
constant poses a theoretical  challenge however and no compelling argument for it 
has been formulated to date. 
Dynamical models for dark energy have also been proposed: quintessence, 
K-essence, fermion condensates, phantom, tachyonic, coupled dark matter and dark 
energy models, to name a few (see \cite{Frieman:2008sn, Sapone:2010iz} for 
recent reviews). 
Quintessence models introduce scalar fields and postulate specific potentials that can 
track dark matter and lead to late time accelerated expansion.  
K-essence models, originally proposed in the context of inflation, modify the kinetic terms 
of scalar fields and also lead to accelerated expansion. For some time period the equation 
of state of the modified scalar field is the one expected for dark energy. 
In another class of models, the (bosonic) long wavelength excitations of a fermion 
condensate play the role of a dark energy field. These excitations are self-interacting 
and at late times relax to an effective cosmological constant. 
In phantom field models, a scalar field exhibits a "wrong sign" kinetic term; even if the 
classical dynamics of such a field can be consistent, the fact that the energy of phantom 
fields is  unbounded below makes their vacuum unstable and their quantum theory not 
well defined unless some mechanism generates a lower bound on their energy. 
Coupled dark matter and dark energy models postulate certain interaction between
(dark) matter and the dark energy field and attempt to explain both the observed 
acceleration as well as the observed similarity of the matter and dark energy densities
today.
Other possible explanations of late-time acceleration, not relying on additional (fundamental 
or effective) scalar fields have also been proposed | such as infrared modifications of general 
relativity. 

All these scenarios, proposed as a consequence of the experimental evidence for dark energy, 
can be made consistent with observations and constraints; their place in a more complete 
theory describing both the evolution of the universe as well as the known particle physics, 
remains to be clarified. 
The Standard Model of particle physics explains successfully all collider and other 
experimental data; it nevertheless has many theoretical shortcomings and it is presently 
seen only as a good effective theory that requires  a high energy completion. Many 
possible extensions of the Standard Model have been proposed, addressing specific 
difficulties of the Standard Model. Some of these models, 
while motivated by particle physics issues, provide good candidates for dark matter.
It seems natural to expect that a compelling solution to both the cosmological observations
and particle physics issues should  exist in the context of a more fundamental theory.

One of the proposed ways for embedding the Standard Model  in string theory is 
by making it part of a quiver gauge theory -- {\it i.e.} a theory with many gauge 
groups and fields transforming either in the adjoint representation
or in the bi-fundamental representation of these groups. Such theories may be realized 
(either directly or holographically) \cite{Verlinde:2005jr}
in terms of D-branes placed at various types of singular points.
The additional gauge groups are either broken at a sufficiently high scale or are 
decoupled from the Standard Model by appropriate choices for the relevant coupling 
constants or other parameters concerning the details of the compactification to four 
dimensions. 
Such decoupled sectors nevertheless affect the observable physics through gravitational 
interactions | either through higher-dimension Plank-suppressed operators or by playing 
the role of dark matter (if the excitations of these sectors are sufficiently massive) or dark 
energy (if the excitations of these sectors are sufficiently light or even massless). 
It is  therefore interesting to study such decoupled theories from this perspective. All scalar 
fields present in these theories transform in nontrivial representations of the gauge group
making them not directly suitable for dark matter or dark energy candidates.
In this paper we will discuss a class of quantum field theories which may appear in 
such a scenario. They are conformally invariant at the classical level but no longer so 
at the quantum level. Conformal invariance will be broken in an interesting way which 
will allow us to gain all-order information about the theory and identify the relevant 
gauge singlet(s) that can play the role of a dark energy field.

Further motivation for analyzing theories of the type outlined above is the proposal 
\cite{FramptonVafa} to embed the Standard Model itself in a conformal field theory. In 
this scenario, at some (high) energy scale particle physics is governed by a 
non-supersymmetric conformal  field theory. Sometimes such theories have unexpected 
features (see {\it e.g.} \cite{Dymarsky:2005uh}, \cite{Dymarsky:2005nc}) whose 
phenomenological consequences are worthwhile exploring.

From a string theory perspective  these theories may be realized in terms of a stack 
of D3 branes probing non-supersymmetric $A_{n-1}$ orbifold singularities; they are 
quiver gauge theories with $n$ nodes. Fields transforming in the adjoint representation 
are realized by strings connecting branes in the same stack while fields in the 
bifundamental representation are the lowest lying modes of strings connecting the 
images under the action of the orbifold group. One may extend such construction 
by adding further D7 branes; apart from matter in the fundamental representation
of the quiver gauge group (which is realized by D3-D7 strings), such a construction
naturally adds gravitationally-coupled gauge singlets. Regardless whether the quiver 
gauge theory is interpreted as a hidden sector (as we do in this paper) or if it is 
broken to the Standard Model, these additional fields may be suitable dark matter 
candidates.

In the next section we will briefly describe a class of non-supersymmetric models in 
which conformal symmetry is broken dynamically; 
due to the special properties of their renormalization group flow, these theories 
can reach a regime in which they simplify dramatically, their entire dynamics being 
governed by a simple matrix scalar quantum field theory. In \S3 we will describe a 
curved space version of this model and construct an equivalent model whose fields are 
singlets under the gauge symmetry of the original model and thus may play the role of a  
dark energy field. 
The cosmological implications of this model will be discussed in \S4. We present 
our conclusions and further comments in \S5.

\section{Flat space models}

Quiver gauge theories have long been subject of 
active investigation.  If the gauge groups are of $SU(N_i)$ type, the action
possesses a variety of double-trace terms required by the absence of the $U(1)$
factors. These double-trace terms renormalize independently of the single-trace 
part of the action. If the coupling constant of the single-trace part of the action 
is fixed, the double-trace coupling runs without bound at low energies implying 
that, in some regime, the double-trace operators dominate the dynamics of the theory.

This scenario is realized in certain nonsupersymmetric theories in the multi-color limit, 
such as orbifold theories  and non-supersymmetric $\beta$-deformed theories
\cite{Dymarsky:2005uh}, \cite{Dymarsky:2005nc}, \cite{up}. The former theories, which will be our main interest in the following, is obtained from
\be
\label{neq4action}
{\cal L}_0 = -\frac{1}{2}(|\Gamma|N)\,\Tr\Big[\; F_{\mu\nu}F^{\mu\nu}
   +D_\mu\Phi^ID^\mu\Phi_I +\lambda^2[\Phi^I,\Phi^J][\Phi_I,\Phi_J]
   +{\bar\chi}D\llap/\chi+\lambda{\bar\chi}\gamma^I[\Phi_I,\chi]\;\Big]
\ee
by an orbifold projection by a discrete group $\Gamma\subset SU(4)$. Here the gauge 
group is $SU(|\Gamma|N)$, the trace is normalized such that the identity element of the 
gauge group has trace equal to $|\Gamma|N$, $\lambda=g_{YM}^2|\Gamma|N$ 
is the 't~Hooft coupling constant, $\Phi^I$ are six real scalar fields transforming in the 
vector representation of $SO(6)$ and $\gamma^I$ are the $SO(6)$ Dirac matrices. As usual, covariant derivatives describe the coupling with the gauge field and are proportional to a single power of the coupling constant $g_{YM}$.
Four fermions, not shown in equation~(\ref{neq4action}), 
transforming in the spinor representation of $SO(6)$ and having Yukawa
couplings with the scalars, complete the action. 
We will denote by $g$ the representation of the elements of $\Gamma$ in 
$SU(|\Gamma|N)$, where they act by conjugation and by $r_g$ and
$R_{g}$ the representation of $\Gamma$ in the spinor and 
vector representation of $SO(6)$, respectively.
$R_g$ and $r_g$ are not unrelated; if one presents the vector representation of 
$SO(6)$ as the 2-index antisymmetric tensor representation of $SU(4)$, then $R_g=r_g\otimes r_g$. The orbifold projection retains in the action (\ref{neq4action}) only 
the components of the original fields obeying the following relations:
\begin{eqnarray}
\label{projection}
A_\mu=g\,A_\mu\,g^\dagger
~~~~~~~~
\phi^I=R_g^{IJ}\,g\,\phi^J\,g^\dagger
~~~~~~~~
\chi^i=r_g^{i}{}_j\,g\,\chi^j\,g^\dagger\,.
\end{eqnarray}
Moreover, the overall factor of $\Gamma$ disappears and the 't~Hooft coupling of the orbifolded theory is $\lambda = g_{YM}^2N$.

Perhaps the simplest example of this construction is for $\Gamma=\IZ_2$ ({\it i.e.} 
$|\Gamma|=2$); its nontrivial element acts as
\be
g={\rm diag}({\bf 1}_N,-{\bf 1}_N)
~~~~~~~~
r_g=-{\bf 1}_4
~~~~~~~~
R_g = {\bf 1}_6
\ee
inside the original $SU(2N)$ gauge group, and on the spinor and vector 
representations of $SO(6)$, respectively. The resulting theory \cite{KT} contains 
two $SU(N)$ gauge groups, 6 scalar fields transforming in the adjoint representation
of the first group, 
6 scalar fields transforming in the adjoint representation
of the second group, 4 fermions transforming in the bifundamental 
$({\bf N},{\bar {\bf N}})$ and 4 fermions transforming in the bifundamental 
$({\bar {\bf N}}, {\bf N})$.

The single-trace part of such orbifold theories has been analyzed in detail in 
\cite{BKV, BJ} where it was shown that these terms are inherited from the parent 
theory; in the case of the action (\ref{neq4action}) (together with its fermionic completion)
they are finite.  
As we previously mentioned however, the action contains additional double-trace terms
\cite{Dymarsky:2005uh, Dymarsky:2005nc} which receive nontrivial infinite 
renormalization. For the simple $\Gamma=\IZ_2$ example they are
\be
\delta {\cal L}_0 = -f_{20}O^{IJ}O_{IJ}-f_0 O^2
\label{egdouble}
\ee
with
\be
O^{IJ}=\Tr[g \Phi^I\Phi^J]- \frac{\delta^{IJ}}{6}\Tr[g \Phi^K\Phi_K]
~~~~~~~~
O=\Tr[g \Phi^I\Phi_I]~~.
\ee

Such double-trace terms are generic in orbifold field theories and may always be traced to 
certain twisted auxiliary fields; they are not
protected from renormalization by the general arguments of \cite{BKV, BJ}.
As shown in \cite{Dymarsky:2005uh}, they are in fact required by the 
renormalizability of the theory. The coupling constant $f$ of a double-trace operator 
$|O|^2$ runs at 1-loop as 
\begin{equation} \label{genbeta} 
M {\partial f\over \partial M} = \beta_f
=  v_O f^2 + 2 \gamma_O \lambda f+ a_O \lambda^2
~~ ,
\end{equation}
where $\lambda=g_{YM}^2N$ is the 't~Hooft coupling, $v_O$ is the 2-point function 
coefficient of the operator $O$, $\gamma_O$ is the anomalous dimension of $O$ and 
$a_O$ is the contribution of the single-trace Lagrangian to the $\beta$-function of the 
double-trace coupling. If the discriminant
\be
D=\gamma^2_O-4 v_O a_O>0
\ee
then $\beta_f$ has a nontrivial zero and $f$ will flow to it.  Some supersymmetry appears
to be required \cite{Dymarsky:2005uh} for such nontrivial zeroes.
If supersymmetry is completely broken, as in the $\IZ_2$ model described above, there 
exists at least one double-trace coupling whose discriminant $D$ is negative, leading to 
the following running coupling:
\begin{equation} \label{expsol}
f(M) = - {\gamma_O\lambda\over v_O} +
{b\lambda\over v_O} \tan \left (
{b\lambda\over v_O}\; \ln\left( \frac{M}{\mu_r} \right)\right )
\ .
\end{equation}
Here we defined $b=\sqrt{-D}$ and, for simplicity, the chosen boundary 
condition is $f(\mu_r)=-d_O\lambda/v_O$. 
Thus, at weak `t Hooft coupling
$\lambda$, the double-trace parameter $f$ varies very slowly
for a wide range of scales. Still, it blows up
towards positive infinity in the UV at
$M = \mu_r e^{\pi v_0/(b\lambda)}$ and reaches $-\infty$ in the IR
at
$M = \mu_r e^{-\pi v_0/(b\lambda)}$. \footnote{One may expect that 
this singular behavior is softened by the $1/N$ corrections, which 
introduce a positive beta function for $\lambda$, making it approach
zero in the IR. We will restrict ourselves to the large $N$ limit.} 
This unexpected runaway behavior of the coupling $f$ was interpreted 
\cite{Dymarsky:2005uh} in terms of a tachyonic instability of the string 
theory dual. This is however not problematic as it has been argued from 
several different standpoints \cite{up, Pomoni:2008de,Horowitz:2007pr} 
that the energy dynamically becomes bounded from below and the theory 
may flow to a nontrivial IR fixed point.

The expectation following from this analysis is that along this flow, the twisted 
dimension-2 single-trace operators appearing in the double-trace operators that are 
generated quantum mechanically as illustrated in eq.~(\ref{egdouble}) acquire 
nontrivial vacuum expectation values. 
To capture this effect and describe the consequences of this
process it is therefore necessary to evaluate the effective potential for these operators; 
from the standpoint of the orbifold theory this is a multi-trace effective potential, the 
dimension-2n term having n traces.
The explicit calculations of \cite{Dymarsky:2005uh} show that the only effect of 
1-loop corrections in the orbifold action without double-trace operators is to 
generate double-trace operators. All other possible terms are forbidden either by the 
analysis of \cite{BKV, BJ} or are derivative terms.  The addition of double-trace 
operators to the tree-level action renders the theory renormalizable.
It is not difficult to see that in this deformed theory, all n-trace terms generated 
at 1-loop level receive contributions {\it solely} from the double-trace deformation 
for all $n\le 3$. Thus, for the purpose of finding the multi-trace effective potential.  
we may dispense with the large set of fields of the orbifold action and capture the 
dynamics of scalar fields by considering a much simpler action -- that of a purely 
bosonic matrix scalar field theory with a specific quartic double-trace  potential. 
To capture the effects of the ignored gluons and fermions all one needs to do is to 
renormalize the divergences and use as running coupling that of the orbifold 
theory (\ref{expsol}). It is important to notice that, as a twisted operator develops 
a vacuum expectation value, the potential of the undeformed orbifold theory becomes 
nonvanishing as well. This constant, which is positive and of the order of the square 
of the vacuum expectation value of the twisted operator, should also be added to 
the effective potential.

While in general there will be several double-trace terms that have the features described 
above, one may focus on the one whose coupling constant runs fastest. This is the case 
because, as indicated by eq.~(\ref{expsol}), the runaway behavior -- and therefore the 
appearance of  nontrivial vacuum expectation value for a twisted operator -- sets in first for 
this operator. Including the other double-trace terms may be treated as a perturbation and 
the running of their coupling constant will likely be affected by the emerging vacuum 
expectation value of the leading twisted operator.
%
It is not always clear which orbifold group element generates the dominant double-trace 
term; we will generically denote this element by $\gamma\in\Gamma\subset 
SU(|\Gamma|N)$ as we will not need its detailed properties. It is however 
typically the case 
\cite{Dymarsky:2005nc}  that the operator whose coupling constant runs fastest is 
constructed out of $SO(6)$ singlets.
Thus, the action we will be interested in is just
\be
{\cal L}= - N\Tr[\eta^{\mu\nu}(\partial_\mu{\bar\Phi})(\partial_\nu\Phi)] 
- f (M)\,|\Tr[\gamma\Phi\bar\Phi]|^2 
\label{ini}
\ee
where without loss of generality we kept only two of the six scalar fields and have 
organized them into complex combinations.

While simpler than the original orbifold action, the action (\ref{ini}) is still complicated 
mainly due to the existence of the matrix degrees of freedom. Its vacuum 
structure and, as we will see in the next section, its gravitational effects, may 
be described in terms of a simpler theory containing a single scalar field to which 
it is completely equivalent.  This is a standard approach: we introduce an auxiliary 
field that linearizes the quartic interaction 
\be
L=-N\eta^{\mu\nu}\Tr[\partial_\mu{\bar\Phi}\partial_\nu\Phi]  
+\frac{1}{f(M)}{\bar\varphi}\varphi
-\varphi \Tr[\gamma^{-1}\Phi{\bar\Phi}]
- {\bar\varphi} \Tr[\gamma\Phi{\bar\Phi}]
\label{waux}
\ee
and construct the 1-particle-irreducible effective action of this field. This action 
contains the complete quantum information about the original theory regardless
of the size of the expectation value of the auxiliary field $\varphi$ (it is, in fact 
not clear a priori that such an expectation value is generated at all)\footnote{It should 
be mentioned that for the toy model in (\ref{ini}) the 1PI effective action is 1-loop exact. 
This is, however, not the case for the orbifold theory which is the motivation for this 
discussion and therefore we will not use this  fact in the following.}. 
%
%
All higher-point potential interactions of the original scalar field $\Phi$ that 
are generated at the quantum level are encoded in the higher-point interactions of 
the auxiliary field. This construction replaces the scalar field $\Phi$ carrying gauge 
degrees of freedom by a single gauge-invariant effective field which may in principle 
be observable, albeit only gravitationally coupled with the Standard Model in the setup 
discussed here.


The scalar theory (\ref{ini}) encodes, through the scale dependence of the coupling 
constant $f(M)$, the effects of the truncated fields on the dynamics of the scalar field
appearing in the twisted operator. The reverse effects -- {\it i.e.} the effects of the 
scalar field evolution on the gauge fields, fermions and the other scalars -- are ignored.
At sufficiently low energies, when the vacuum expectation value of the twisted operator
(or, alternatively, for the auxiliary field $\varphi$) becomes comparable to the energy 
scale, this back-reaction can no longer be ignored. 
Accounting for it is crucial for the arguments \cite{up, Pomoni:2008de,Horowitz:2007pr} 
that the theory flows to a nontrivial IR fixed point. 
Heuristically, around this energy scale, parts of the gauge fields, fermions and scalars 
become sufficiently massive and no longer contribute to the RG coefficients and a nontrivial 
beta function for the gauge coupling is generated.
Consequently, the running of the double-trace coupling (\ref{genbeta}) will 
be supplemented by the running of the gauge coupling and the resulting system is expected 
to no longer have a runaway solution. 
On symmetry grounds, the fixed point theory is not expected to depend on the vacuum 
expectation value of the twisted operators, which is therefore irrelevant in a flat space setup. 
In curved space however, the vacuum expectation value sources gravitational field and
its gravitational effects  will not change qualitatively along the RG flow of the complete theory.


In general, the orbifold field theory (\ref{neq4action})-(\ref{projection}) as well as 
the reduced model 
(\ref{ini}) are symmetric under the discrete transformation 
$\gamma\rightarrow \omega \gamma$ where $\omega$ is a root of unity of 
the appropriate order to keep $\omega \gamma$ an element of the orbifold group. 
Introduction of the auxiliary field does not break this symmetry which is preserved 
if the auxiliary field transforms as $\varphi\rightarrow \omega\varphi$, as may be 
readily seen from equation (\ref{waux}).  This symmetry constrains the 1PI effective 
action  to be solely a function of $\varphi{\bar\varphi}$; in particular, tadpole-like 
terms linear in $\varphi$ or ${\bar\varphi}$ are forbidden.

The potential for the auxiliary field typically has a nontrivial minimum \cite{up}, 
which translates into a nontrivial vacuum expectation value for the twisted operator 
$\Tr[\gamma\Phi{\bar\Phi}]$. This can be translated into a nontrivial expectation 
value for the scalar fields. Such expectation values have two distinct effects: (a) they 
may yield a nonzero positive value for the tree-level potential and 
(b) at 1-loop they contribute a positive constant term to the potential. In flat space 
both contributions are, of course, irrelevant. The details of these contributions 
depend on the details of the orbifold group. 

In the next section we will place the model discussed here in curved space. 
One may find that a finite density of $\Phi$ quanta is created (in a 
gauge-invariant configurations); to account for this possibility we also add a 
chemical potential. A systematic way of accounting for this  
is to add the number operator to the Hamiltonian which is then transformed to a 
Lagrangian framework.
Since the auxiliary field $\varphi$ linearized the quartic interaction, the behavior of the
resulting field theory in the presence of a chemical potential is quite analogous to the zero 
temperature limit of an ideal relativistic Bose gas. Classically, the fields $\Phi$ are 
massless with the interaction terms playing the role of an effective mass. 
At the quantum level the auxiliary fields ${\varphi}$ and ${\bar\varphi}$ acquire 
nontrivial vacuum expectation values, justifying this interpretation.

Unlike systems of fermions, in the zero temperature limit the chemical potential does not 
enter explicitly the expression of the bosonic partition function. 
Rather, it is determined separately
from the requirement that all particles are found in the ground state. 
If $\epsilon_0$ is the ground state energy (i.e. the effective mass of the $\Phi$ particles) 
then the chemical potential is determined by the particle number density in the ground 
state: \footnote{\label{longfootnote}
In general it is not possible to give different interpretations to the modes 
of a field depending on their energy. However, modes with specified spatial momenta of any 
one field are a set of measure zero in the path integral. Due to the contribution 
of the integration measure, $\int dE \int k^2 dk,$ modes with vanishing spatial 
momenta make vanishing contributions to the 1-loop grand potential in the continuum
limit. In a theory placed in finite volume one may isolate the contribution of the
lowest energy mode. Its contribution is volume independent and thus subleading 
for large/infinite spatial volume which is the case for an FRW universe or in 
Minkowski space.
f
o
This is a standard approach which is used extensively in the treatment of Bose 
condensation; its advantage is that it allows particles to dynamically condense to the 
ground state or leave the condensate. However, the ground state condensate should not, in 
our case, be described as a classical field. From equation (\ref{waux}) it is clear that the 
auxiliary field $\varphi$ acts as effective mass for the fields $\Phi$. As we will see in 
later sections, the auxiliary field acquires generically a nontrivial expectation value thus 
rendering $\Phi$ effectively massive. This effective mass is the ground state energy. Any 
translational momenta for the ground state quanta will raise their energy and this take
them out of the ground state. 
We should also note that, since they have vanishing momenta, {\it in flat space} the ground 
state particles do not contribute directly to the effective potential for the auxiliary field, 
their contribution being cancelled by the integration measure.
}
\be
n=\frac{1}{e^{\beta(\epsilon_0-\mu)}-1}
~~~\Rightarrow~~~
\mu=\epsilon_0+\frac{kT}{n}\simeq \epsilon_0\,.
\ee
The standard relation between the grand potential and the partition function yields then 
the energy density as
\be
\rho = \lim_{T\rightarrow 0}(-kT\ln Z) +\mu n \,.
\label{Edensity}
\ee
One can easily see that the first term is nothing but the zero temperature field theory 
partition function to be obtained from the action (\ref{ini}). It is also the $T_{00}$ 
component of the auxiliary field stress tensor constructed from the quantum effective 
action. Thus, a simple way to include a finite $\Phi$ density at zero temperature is to 
add a pressureless stress tensor with $T_{00}=\mu n$ to the stress tensor of 
the auxiliary field $\varphi$. 

Clearly, such an addition manifestly breaks conformal invariance. From the standpoint 
of the orbifold theory, for which (\ref{ini}) is only a model capturing its essentia features,  
all violations of conformal invariance occur only quantum mechanically and are 
proportional to the beta-function of the double-trace operator (in a cosmological setting
further effects proportional to the Hubble constant may also occur). For this reason 
one may expect that the particle-number density $n$ is small; while we keep it throughout
the general analysis, we will set it to zero when exploring the cosmological consequences 
of our model.


In the next section we will discuss the curved space version of the quantum field theory 
detailed above and compute its partition function. It will turn out that, with a suitable 
choice of variables, the curved space calculation can be easily mapped to a flat space 
calculation.

\section{Curved space model}

The effective model we identified in the previous section captures the features of a 
quantum field theory whose only departure from scale invariance is encoded in the 
running of a certain double-trace coupling constant $f$. In coupling it to gravity 
we would like to preserve this feature; to this end, apart from the standard 
covariantization of index contraction, we will also add a conformal coupling, $\xi=1/6$, 
for the scalar fields $\Phi$.\footnote{Such a coupling is required for the renormalizability 
of the theory in curved background; the value $\xi=1/6$ ensures scale invariance.}
\be
L=\sqrt{-g} \left(- Ng^{\mu\nu}\Tr[\partial_\mu{\bar\Phi}\partial_\nu\Phi]  
                     - \xi \,R\, \Tr[\bar\Phi\Phi]- f |\Tr[\gamma\Phi\bar\Phi]|^2 
\right)
\label{iniScalar}
\ee
where, as before, $\gamma$ is an element of the orbifold group 
$\Gamma\subset SU(N)$ and $R$ is the Ricci scalar.  

Similarly to the flat space theory, we proceed by linearizing the
dependence on $\Phi$ by introducing auxiliary fields:
\be
L=\sqrt{-g} \left(-Ng^{\mu\nu}\Tr[\partial_\mu{\bar\Phi}\partial_\nu\Phi]  - \xi R \Tr[\bar\Phi\Phi] 
+\frac{1}{f}{\bar\varphi}\varphi
-\varphi \Tr[\gamma^{-1}\Phi{\bar\Phi}]
- {\bar\varphi} \Tr[\gamma\Phi{\bar\Phi}]\right)~.
\ee
Scale invariance implies that a rescaling of the metric $g_{\mu\nu}$ can be absorbed 
by a field redefinition. In conformal time, an FRW metric is just that of Minkowski space
up to a scale factor $a(\eta)$. The presence of the conformal coupling then implies  that
this scale factor can be eliminated by a field redefinition. Indeed, it is easy to see that
in terms of the new fields 
\be
{\hat \Phi}={a(\eta)}\,\Phi~~~~~{\hat\varphi}={a(\eta)^2}\,\varphi
\ee
the action is exactly that of the original Minkowsky space theory:
%
\be
L=-N\Tr[\partial_\mu{\bar{\hat\Phi}}\partial^\mu\hat\Phi] 
+\frac{1}{f}{\bar{\hat\varphi}}\hat\varphi
-\hat\varphi \Tr[\gamma^{-1}\hat\Phi{\bar{\hat\Phi}}]
- {\bar{\hat\varphi}} \Tr[\gamma\hat\Phi{\bar{\hat\Phi}}]\,.
\ee
This is a reflection of the scale invariance of (\ref{iniScalar}). 

The scalar fields $\hat\Phi$ appear quadratically. Therefore the 1PI effective potential may 
be easily computed by evaluating the determinant of the operator
\be
K_B = -\eta^{\mu\nu}\partial_\mu\partial_\nu
+\frac{1}{N}\hat\varphi\,\gamma^{-1}
+\frac{1}{N}{\bar {\hat\varphi}\gamma}~~,~~~~~V_{\text{eff}}&=& i \ln\det K_B ~.
\ee
On general grounds this determinant has both power-like and logarithmic UV divergences. 
These divergences may however be traced to the fact that we are focusing on the dynamics of 
the fields $\Phi$ and are ignoring the other fields. 
%
If a quadratic divergence were present, it would generate a tadpole for $\hat\varphi$;
in the complete orbifold theory such a term would suggest a perturbative generation 
of a mass term for some component of $\Phi$ at two loops. This however is not expected 
to happen in the large $N$ limit \cite{BKV, BJ}, consistent with the cancellation of terms 
linear in $\hat\varphi$. We will therefore discard quadratic divergences from our effective model.

With this clarifications, and if there exists a choice of fields $\hat\varphi$ and 
analytic continuation of momenta such that $K_B$ does not have zero eigenvalues,  
the loop-induced potential for the auxiliary fields $\hat\varphi$ and  
$\bar{\hat\varphi}$ is just
\be
V_{\text{eff}}
&=& \frac{1}{(4\pi)^2}\Tr\left[
{\cal M}^2\left(
\ln\frac{{\cal M}^2}{\Lambda^4}-1\right)\right]
\label{CW}
\ee
with
\be
{\cal M}=\frac{1}{N}\left({\hat\varphi}\,\gamma^{-1}+{\bar {\hat\varphi}\gamma}\right)
~.
\ee
Here we carried out the trace over momenta; the remaining trace is only over the 
gauge group indices. In the complete theory the logarithmic divergence present in equation (\ref{CW}) contributes to the running of the coupling constant $f$ through the 
coefficient $v_O$ in the $\beta$-function (\ref{genbeta}). After renormalization 
the cutoff scale $\Lambda$ is replaced with the renormalization scale $\mu_r$.

For a $\IZ_n$ orbifold, the Coleman-Weinberg potential (\ref{CW}) is:
\be
V_{\text{eff}}&=& 
 \frac{N^2}{(4\pi)^2}\,\sum_{k=0}^n\left[m_k^2\left(
\ln\frac{m_k^2}{{\mu_r}^4}-1\right)\right]\nonumber
\\
m_k&=&\frac{1}{N}\alpha_k\hat\varphi+\frac{1}{N}\alpha_k^*{\bar{\hat\varphi}}~~~~~~~~\alpha_k^n=1
~~~~\alpha_k\alpha_k^*=1~,
\ee

In the calculation above ${\hat \varphi}$ maintains its role as a non-dynamical field as
it was assumed to be constant throughout the calculation. It is however easy to 
see that derivative terms are also generated at the quantum level. They may be 
organized following the number of fields on which the derivatives act. To this end 
one separates ${\cal M}$ into a constant part and a position/momentum-dependent part
and expands in the latter. The terms in which no derivatives act on the 
momentum-dependent part of ${\cal M}$ may be resummed and lead to a trivial shift
of the constant part of ${\cal M}$.
This is similar to doing perturbation theory around an arbitrary value of $\hat\varphi$. 
While in general this would manifest an instability, this is not the case here because
at tree level any value of $\hat\varphi$ is allowed (alternatively, because the potential for 
$\hat\varphi$ is generated at the same order as the derivative terms).
With the same assumption as before, that momenta may be analytically continued  
such that for fixed fields $K_B$ does not have zero eigenvalues, 
the first correction, containing two derivatives, is given by\footnote{The evaluation 
of this correction
relies on the observation that $[{\cal M}(0),{\cal M}(q)]=0$. This holds as both
${\cal M}(0)$ and ${\cal M}(q)$ are constructed form mutually commuting matrices.}
\be
\delta L_K = \frac{1}{2}\Tr
\left[
\frac{1}{p^2+{\cal M}}{\cal M}(-q)\frac{1}{(p+q)^2+{\cal M}}{\cal M}(q)
\right]-\frac{1}{2}\Tr
\left[
\frac{1}{(p^2+{\cal M})^2}\right]\Tr[{\cal M}(-q){\cal M}(q)]~.
\label{derivativeterm}
\ee
Certain care is necessary in the identification of the leading term in the momentum 
expansion, which depends on the details of the orbifold group and choice of group 
element $\gamma$.
Regardless of its precise expression, the meaning of  this term from the standpoint 
of the original theory is that of an effective contribution to the 4-point scalar amplitude.
Following a strategy similar to the above it is not difficult to find higher-derivative 
corrections. We will however refrain from writing general expressions here.

\subsection{An illustrative example}

The simplest example illustrating the discussion in the previous section is the 
$\IZ_2$ orbifold theory, {\it i.e.} $n=2$; $m_0=-m_1$. The corresponding numerical 
coefficients $\alpha_k$ are such that $\alpha_0^2=1=\alpha_1^2$. In this case the 
field $\hat\varphi$ may be chosen to be real. Repeating the discussion above and 
accounting for the fact that in this case ${\cal M}$ always has a negative eigenvalue,
we find that the effective potential is given by
\be
V_{\text{eff}}^{\mathbb{Z}_2}&=& 
+\frac{8\, {\hat\phi} ^2}{(4\pi)^2}\left(
\ln\frac{4\hat\phi{}^2}{N^2\,\mu_r^4}-1\right)~.
\ee
The complete potential is therefore
\be
\label{fullV}
V^{\mathbb{Z}_2}=\frac{c_0}{f}{\hat\varphi}_0^2
-\frac{\hat\varphi^2}{f} + V_{\text{eff}}^{\mathbb{Z}_2}\,.
\ee
As discussed in the previous section between equations (\ref{expsol}) 
and (\ref{ini}), we have added to the effective potential the effects of the 
scalar potential of the undeformed orbifold theory at the critical points 
$\hat\varphi_0$  of $V_{\text{eff}}$. As discussed there, it is a constant 
of the order of the square of the vacuum expectation value of the double-trace 
operator -- {\it i.e.} it must be proportional to $(\hat\varphi_0)^2/f$. $c_0$ is a 
numerical coefficient of order unity.
i

For this simple case 
the two-derivative term (\ref{derivativeterm}) is just:
%
%
\be
\delta L_K =-\frac{N}{6(4\pi)^2}\;{\hat\varphi}^{-1}\,
\partial^\mu{\hat\varphi}\partial_\mu{\hat\varphi}~.
\ee
This will promote the auxiliary field $\hat\varphi$ to a dynamical field. Higher 
derivative terms have the structure
\be
\label{higher_derivatives}
\sum_n \frac{c_n}{\hat\varphi{}^{n+1}}\partial^\mu{\hat\varphi}
\Box^n\partial_\mu{\hat\varphi}
\ee
with numerical coefficients $c_n$; further terms, in which derivatives act on three 
or more fields, are also generated. We will neglect all such terms in the following, 
assuming that ${\hat \varphi}$ varies sufficiently slowly.

To identify the action for the auxiliary field $\varphi$ (and consequently its equation 
of state) it is necessary to restore the background FRW metric and also return to the 
comoving frame. 
It is therefore important to understand the behavior of the renormalization 
scale $\mu_r^2$ under this transformation. 
In curved space any cutoff or other scales should have a covariant interpretation. 
Thus, $\mu_r^2$ should be thought of as a square taken with the (inverse) 
metric; this implies that, as we restore the background FRW metric, we should 
also replace
\be
\mu_r^2\rightarrow a(\eta)^2\mu_r^2~~.
\ee
The potential therefore becomes
\be
V^{\mathbb{Z}_2}=a(\eta)^4\left[
\frac{c_0}{f}\varphi_0^2
-\frac{\varphi^2}{f} +\frac{8}{(4\pi)^2}\,\varphi^2\left(
\ln\frac{4a(\eta)^4\varphi^2}{a(\eta)^4N^2\mu_r^4}-1\right)\right]
=\sqrt{-g}\,V(\varphi)
\ee
Thus, after identifying covariant quantities, the potential for $\varphi$ is
\be
V(\varphi) = 
\frac{c_0}{f}\varphi_0^2
-\frac{\varphi^2}{f} +\frac{8}{(4\pi)^2}\,\varphi^2\,\left(
\ln\frac{4\varphi^2}{N^2\mu_r^4}-1\right)\, .
\ee
Restoring the scale factor in the derivative terms also leads to the appearance of 
its derivatives (and thus of powers of curvature invariants) and of additional 
potential-like terms. They turn out to have simple covariant
expressions, being expressible in terms of the Ricci scalar: 
\be
\delta L_K &=&-\frac{N}{6(4\pi)^2}\,\sqrt{-g}\;\frac{1}{a(\eta)^4\varphi}\,
g^{\mu\nu}\partial_\mu(a(\eta)^2\varphi)\partial_\nu(a(\eta)^2\varphi)
\\
&=&-\frac{N}{6(4\pi)^2}\,\;\sqrt{-g}\left[
 \frac{1}{\varphi} g^{\mu\nu}\partial_\mu\varphi\partial_\nu\varphi
+\frac{2}{3}\varphi\,R\right]~.
\ee
The power of $\varphi$ multiplying the Ricci scalar is determined by dimensional 
analysis; the appearance of the last term is a direct consequence of the conformal 
invariance of the original model.

The derivative term may be brought to standard form by a simple field redefinition
%
\be
\varphi = \frac{3}{4N}(4\pi)^2\,\zeta^2~~;
\ee
the new field $\zeta$ has the canonical dimension of a scalar field. The 
complete (two-derivative) Lagrangian for this new field is
\be
\delta L &=&-\frac{1}{2}\,\left[
 g^{\mu\nu}\partial_\mu\zeta\partial_\nu\zeta
+\frac{1}{6}\zeta^2\,R\right]
\\
V(\zeta)&=&
\frac{c_0}{f}\,\varphi_0^2
-\frac{9(4\pi)^4}{16\,N^2\,f}\zeta^4 
+\frac{9(4\pi)^2}{2\,N^2}\,\zeta^4\left(
\ln\frac{9(4\pi)^4\zeta^4}{4\,N^4\,\mu_r^4}-1\right)\\
L_{\text{eff}}&=&\frac{1}{2G}R+\delta L - V
\label{fin_action}
\ee
where $G$ is related to Newton's constant by $G=8\pi G_N$. 
The scale factor may also be restored in the higher-derivative 
terms (\ref{higher_derivatives}). 
The Einstein-Hilbert term can be cast in canonical form by rescaling the metric by 
$(1-G\zeta^2/6)$. 
We will keep the action in its current form.


\section{Cosmological consequences}

In previous sections we argued that, if the Standard Model is embedded in string theory
through a certain class of quiver gauge theories, then decoupled sectors of that theory 
yield effective actions of the type (\ref{fin_action}) which interact only gravitationally 
with the usual matter fields. It is therefore interesting to explore the cosmological 
implications of such actions.  

While different in details, the model constructed above exhibits elements of classes of 
models discussed elsewhere. The existence of a field-dependent Newton's constant 
makes it similar to modified gravity models.\footnote{Let us note here that terms 
containing curvature invariants and matter fields are generic at higher order in 
perturbation theory.} 
As we will see in the following, the same 
effective Newton's constant (or the conformal coupling of the field $\zeta$) will lead to 
the matter energy density acting as a source for $\zeta$ and vice versa\footnote{For
similar reasons, such matter-dark energy couplings are also generic in modified gravity
models.}. 
Finally, the 
existence of nontrivial derivative terms and potential terms for $\zeta$ makes it similar 
to K-essence and quintessence models. While of course the field-dependent Newton's constant 
may be eliminated by a suitable Weyl rescaling of the metric, all the other features of the 
model survive this transformation.  

Let us proceed with analyzing the cosmological implications of the action (\ref{fin_action});
as we will see, we will recover many desirable qualitative properties of cosmological 
parameters. We will also see that the effective cosmological constant will turn out to exhibit 
an exponential suppression compared to the renormalization scale.
It is not difficult to see that the equation of motion for the field $\zeta$ and Einstein's 
equations are:
\be
\label{zetaeq}
\Box\zeta -\frac{1}{6}\zeta R -\frac{\partial V}{\partial\zeta}&=&0
\\
\label{Eeq}
E_{\mu\nu}&=&G\left(T^\zeta_{\mu\nu}
+T^c_{\mu\nu}+\delta T^\zeta_{\mu\nu}+T^m_{\mu\nu}\right)\\
T^\zeta_{\mu\nu}&=&
\partial_\mu\zeta\partial_\nu\zeta-\frac{1}{2}g_{\mu\nu}
\left(g^{\sigma\rho}\partial_\sigma\zeta\partial_\rho\zeta +2V\right)
\\
T^c_{\mu\nu}&=&\frac{1}{6}\zeta^2 E_{\mu\nu}
+\frac{1}{6}\left[g_{\mu\nu}\;\Box(\zeta^2)-\nabla_\mu\nabla_\nu\zeta^2\right]
\\
\label{delTzeta}
\delta T^\zeta_{\mu\nu}
&=&g_{\mu 0}g_{\nu 0} \sqrt{2 x}|\zeta| \frac{n_0}{a(t)^2}
\ee
where $\Box$ stands for the usual covariant Laplacian operator. 
In the equations above $T^\zeta$, $T^c$, $\delta T^\zeta$ and $T^m$ are
the stress tensor of the $\zeta$ field in the absence of the conformal coupling, 
the contribution of the conformal coupling to the stress tensor, the stress 
tensor due to a finite density of $\Phi$ quanta and the matter stress tensor, 
respectively\footnote{The equations (\ref{Eeq})-(\ref{delTzeta}) receive, in principle,
contributions due to particle production due to the expansion of the universe. There is, 
however, no direct contribution to the equation of motion for the dark energy field 
$\zeta$ (\ref{zetaeq}) and therefore the contribution of particle production on the 
evolution of  $\zeta$ is expected to be small. We will not include the effects of particle
production.}$^,$\footnote{In flat space, the vacuum expectation value of the field $\zeta$ 
is the critical point of the effective potential. The curved space analog is the set of 
equations above in which one requires $\zeta=\zeta_0$ to be constant. Unlike the flat 
space case discussed in footnote~\ref{longfootnote}, these equations introduce a 
correlation between $\zeta_0$ and the density $n_0$ of the ground state condensate 
of  $\Phi$ quanta. This is, however, a second order effect which arises due to the gravitational coupling and is therefore expected to be heavily suppressed.}.

The second term in equation~(\ref{zetaeq}) is also a consequence of the conformal 
coupling. It is useful to separate as much as possible the evolution of $\zeta$ from that 
of the scale factor $a$ (or more generally the metric if one were interested in constructing 
the perturbation equations). Evaluating the Ricci scalar from the trace of Einstein's 
equations and replacing it in (\ref{zetaeq}) leads to
%
%
\be
\Box\zeta -\frac{2G}{3}\zeta V 
- \left(1-\frac{G}{6}\zeta^2\right) \frac{\partial V}{\partial\zeta}
+\frac{G}{6}\zeta  g^{\mu\nu }(\delta T_{\mu\nu}^\zeta +T_{\mu\nu}^m)
&=&0~.
\ee
We notice here that both the matter density as well as a finite density of $\Phi$ quanta
source the evolution of $\zeta$. The coupling between them, polynomial in $\zeta$, is 
different from existing analysis of couplings between (dark) matter and dark energy, 
see {\it e.g.} \cite{Amendola:2007yx, Baldi:2010pq, Baldi:2010vv}.

These equations simplify substantially for an isotropic and homogeneous universe 
with flat spatial slices, for which the metric takes the standard FRW form:\footnote{
Absence of a solution with such an ansatz simply means that one or more of the 
assumptions | homogeneity, isotropy or flatness | should be relaxed.}
\be
ds^2 = -dt^2+a(t)^2d{\vec x}^2~.
\ee
The assumptions of isotropy and homogeneity also imply that $\zeta=\zeta(t)$ does 
not have any spatial dependence; then, the equations above simplify to:
\be
\label{zeta}
& & {\ddot\zeta}+3H{\dot\zeta} +\frac{2G}{3}\zeta V 
+ \left(1-\frac{G}{6}\zeta^2\right) \frac{\partial V}{\partial\zeta}
+\frac{G}{6}
\zeta  \left(-\sqrt{2 x}|\zeta| \frac{n_0}{a(t)^2} 
-{\tilde \rho^m}+3 \tilde p^m\right)
=0
\\
& & 
\label{friedman}
3
 H^2=G
  (\rho^\zeta+\rho^m)
\\
& & 
\label{acc}
2\frac{\ddot a}{a}=-\frac{G}{3}
(\rho^\zeta+3p^\zeta+\rho^m+3 p^m)
\ee
where $\rho^m= \frac{\tilde \rho^m}{\left(1-\frac{G}{6}\zeta^2\right) }$
is the effective matter energy density and $p^m$ is the matter pressure, which 
vanishes for non-relativistic matter. 
We have also identified the energy density and pressure of the $\zeta$ fluid as:
\be
\label{rhozeta}
\rho^\zeta&=&\frac{1}{1-\frac{G}{6}\zeta^2}\left(
V+\frac{1}{2}{\dot\zeta}^2+H\zeta{\dot \zeta}+\sqrt{2 x}|\zeta| \frac{n_0}{a(t)^2} \right)
\cr
p^\zeta&=&
\frac{1}{1-\frac{G}{6}\zeta^2}\left(
-V+\frac{1}{6}{\dot\zeta}^2-\frac{1}{3}\zeta\left({\ddot\zeta}+2H{\dot\zeta}\right)\right)
~.
\ee
The equation of state for the $\zeta$ fluid follows immediately:
\be
\label{eqofstate}
1+w=\frac{1}{3}\,\frac{2{\dot\zeta}^2
-\zeta({\ddot\zeta}-H{\dot \zeta})+
12\pi\sqrt{\frac{3}{2N}}|\zeta| \frac{n_0}{a(t)^2}
}{V+\frac{1}{2}{\dot\zeta}^2+H\zeta{\dot \zeta}
+
12\pi\sqrt{\frac{3}{2N}}|\zeta| \frac{n_0}{a(t)^2}
}~.
\ee
At very late times in an expanding universe, if $\zeta$ asymptotes to a constant, it is 
easy to see that the equation above reduces to $w=-1$. In general, we see however that 
the acceleration term $\ddot\zeta$ can help bring the equation of state close to $w=-1$ 
without necessarily requiring a small kinetic energy (or an extremely small mass for the 
$\zeta$ field). This acceleration term is also a consequence of the conformal coupling in
(\ref{fin_action}).

It is curious to notice that the dependence of $\delta T^\zeta$ on the scale factor
 $a$ | identifiable by the coefficient $n_0$ | implies that a nonzero $\Phi$ density acts 
from the standpoint of the  Friedman equation like nontrivial {\it negative} spatial 
curvature. Unlike regular curvature contributions however, this term does not affect 
the curvature of spatial slices and is therefore unconstrained by observations.

In practice it is convenient to replace the acceleration equation with an equation
describing the time evolution of the matter energy density.
As usual, Einstein's equations (\ref{friedman}) and (\ref{acc}) encode the time 
evolution of the total energy density. The time evolution of the energy density of the 
$\zeta$ fluid is, of course, given by equations (\ref{rhozeta}) and (\ref{zeta}). 
It thus follows that
\be
\label{rho}
\dot\rho^m=
-\dot\rho^\zeta-3\frac{\dot a}{a}\left(\rho^\zeta+\rho^m+p^\zeta+p^m\right)~,
\ee
which we will use in place of (\ref{acc}). If $\zeta$ is fixed to a constant this equation 
has the usual solution $\rho^m\propto a^{-3}$. We will see that this behavior naturally occurs at very late times.

\subsection{A special solution}

In general, the equations discussed above yield a nontrivial time-dependent profile for 
the field $\zeta$ which is sourced by the matter density and by the density of $\Phi$ 
quanta populating the ground state. 
As discussed in \S2, the existence of a finite density of  $\Phi$  quanta is a 
quantum mechanical effect (proportional to the $\beta$-function of the 
double-trace operators) and is expected to be small. Thus, assuming no 
accidental enhancements, we will set $n_0=0$.
In the absence of matter and for $n_0=0$ the evolution equations admit a simple 
solution with constant value of $\zeta=\zeta_0$. This solution is deformed nontrivially 
by a nonzero matter density, $n_0\ne 0$ or even by a choice of initial conditions that 
set $\zeta(t=t_0)\ne \zeta_0$ or ${\dot\zeta}(t=t_0) \ne 0$.


Setting $\zeta=\zeta_0$, the equations (\ref{zeta}), (\ref{friedman}) become
\be
\label{zetac}
\frac{2G}{3}\zeta_0 V(\zeta_0) + \left(1-\frac{G}{6}\zeta_0^2\right) 
\frac{\partial V}{\partial\zeta}\Big|_{\zeta=\zeta_0}
&=&0
\\
\label{Hc}
\frac{G\,V(\zeta_0)}{3\left(1-\frac{G}{6}\zeta_0^2\right)}
&=&H^2
\ee
and, as discussed before, the matter density has the usual $1/a(t)^3$ behavior.
The first term in equation (\ref{zetac}) is a consequence of the conformal coupling 
of $\zeta$; in the absence of such a coupling $\zeta_0$ is simply given by the position 
of the minimum of the potential. 
%
The second equation requires that its left hand side be positive.
\footnote{If the left-hand side of this equation were negative, the FRW ansatz for the 
metric no longer yields a solution to Einstein's equations; the appropriate solution 
becomes the anti-de-Sitter space.} The solution to this equation is the standard scale 
factor in the presence of a cosmological constant and an effective Newton's constant 
$G_{\text{eff}}=G/(1-G\zeta_0^2/6)$. As discussed before, the equation of state reduces 
to the usual $w=-1$.

Due to the nature of the equations (\ref{zeta}), (\ref{friedman}) and (\ref{rho}) it is 
natural to expect that, in the far future, any time-dependent profile for $\zeta$ will 
asymptote to the solution described above. It is therefore interesting to discuss it in more 
detail. 
The solution of the counterpart of equation~(\ref{zetac}) in the absence of the 
conformal coupling may be readily obtained:
\be
{\bar \zeta}_0 = \pm \frac{N\,\mu_r}{2\pi\sqrt{6}}\, e^{\pi^2/2f}~.
\ee
We see that it is exponentially smaller than the renormalization scale $\mu_r$ for small 
and negative values of the double-trace coupling $f$. Such values are allowed by 
the special properties of the renormalization group flow discussed in \S2.  As 
mentioned there, $f$ runs very slowly if  the 't~Hooft coupling is small; 
it is therefore natural 
to consider a double-trace coupling that is small (yet larger than the 't~Hooft coupling) 
and fixed over a large range of scales.
With this starting point, the solution to (\ref{zetac}) may be found as a series in 
$G{\bar\zeta}_0$:
\be
\zeta_0={\bar\zeta}_0\left(1+\sum_{n\ge 1}d_n G^n {\bar\zeta}_0^n\right)~~.
\ee
Evaluating the left-hand side of equation (\ref{Hc}) we find that the effective 
cosmological constant is, up to irrelevant numerical coefficients, 
\be
\Lambda_{\text{eff}}=\frac{V(\zeta_0)}{\left(1-\frac{G}{6}\zeta_0^2\right)}
\propto N^4\mu_r^4 e^{-\frac{2\pi^2}{|f|}}\left(1+\sum_{n\ge 1}e_n 
G^n {\bar\zeta}_0^n\right)
\ee
{\it i.e.} it is exponentially small compared to the energy scale $\mu_r$ 
that governs the dynamics of the theory. Clearly, the smallness of the effective 
cosmological constant is a consequence of the exponential factor which is 
independent of the constant $c_0$ in equation (\ref{fullV}), making the 
main qualitative features of our results independent of the precise value of $c_0$. 

\subsection{More general solutions}

While simple to analyze, the solutions discussed above are not generic even in the 
absence of matter and of a finite density of $\Phi$ quanta. \footnote{As in the 
previous section, we will continue to assume that $n_0=0$.}
A typical solution has 
a nontrivial time-dependence triggered by the initial conditions for the $\zeta$ field
which relaxes in the far future to the solution discussed in the previous subsection. 
Perhaps the simplest way to construct such solutions is to decouple the two equations 
(\ref{zeta}) and (\ref{friedman}) by solving for the Hubble parameter from the latter 
and replacing it in the former. The positivity requirement (\ref{Hc}) receives derivative
corrections and becomes
\be
{\dot\zeta}^2\frac{1+\frac{G}{6}\zeta^2}{1-\frac{G}{6}\zeta^2}
+\frac{2V}{1-\frac{G}{6}\zeta^2}\ge 0~.
\ee
While decoupled, the resulting equations are lengthy and nonlinear; finding the 
complete time dependence of their solution requires a numerical approach. 

\begin{figure}
$\!\!\!\!\!\!\!\!\!\!$
\includegraphics[width=8.22cm]{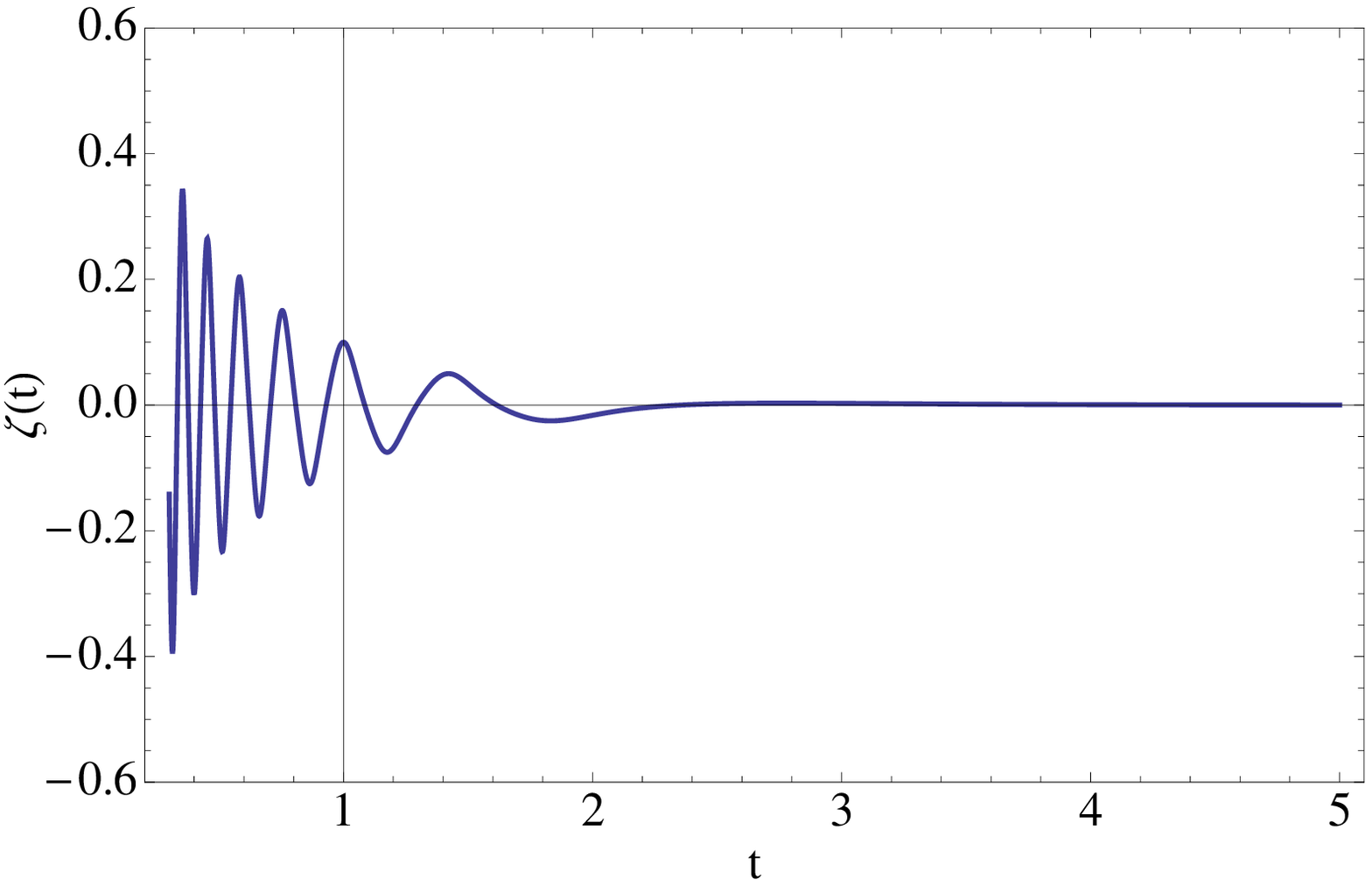}
$\!\!\!\!$
\includegraphics[width=8.22cm]{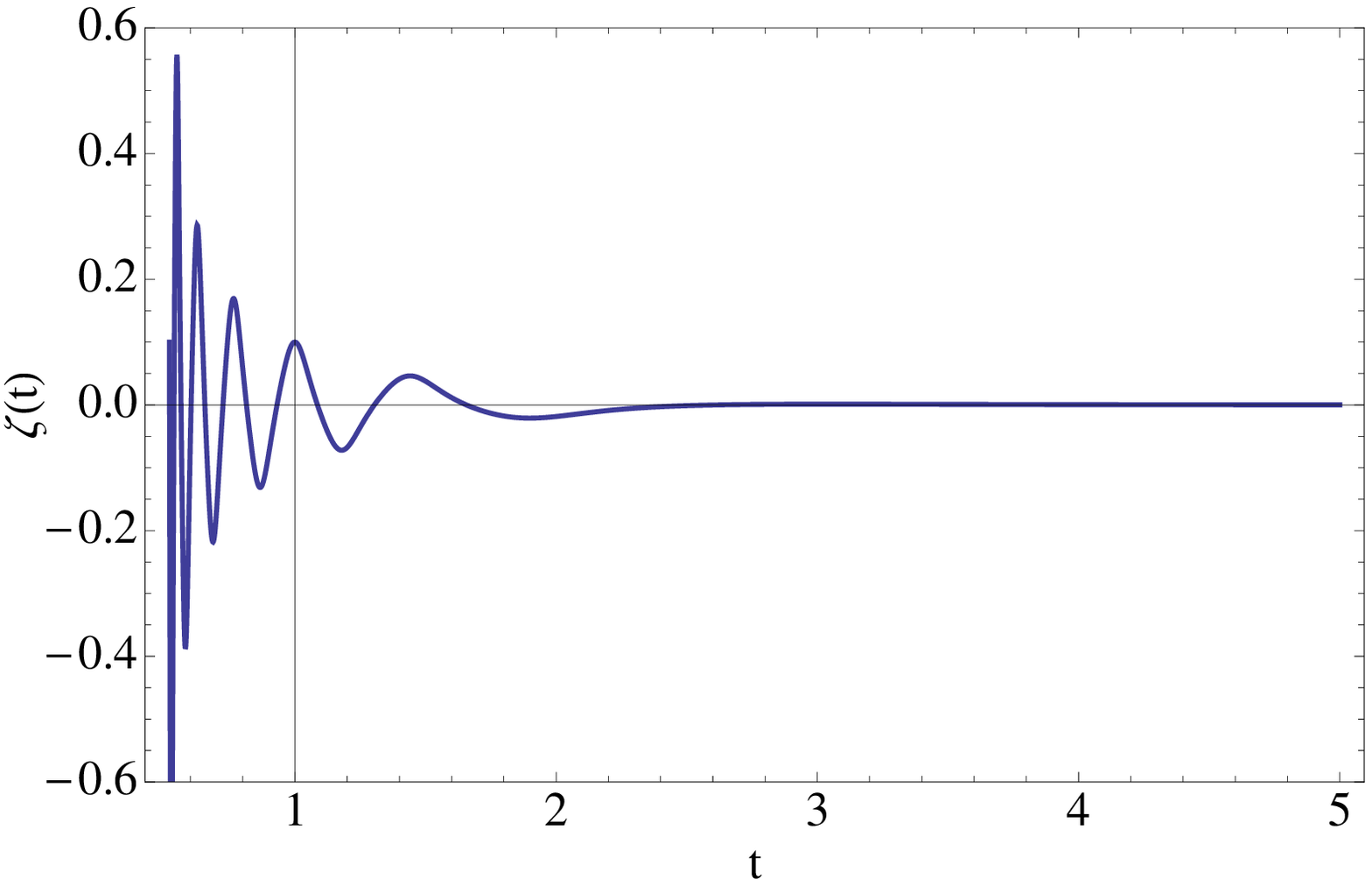}
$\!\!\!\!\!\!$
\includegraphics[width=8.18cm]{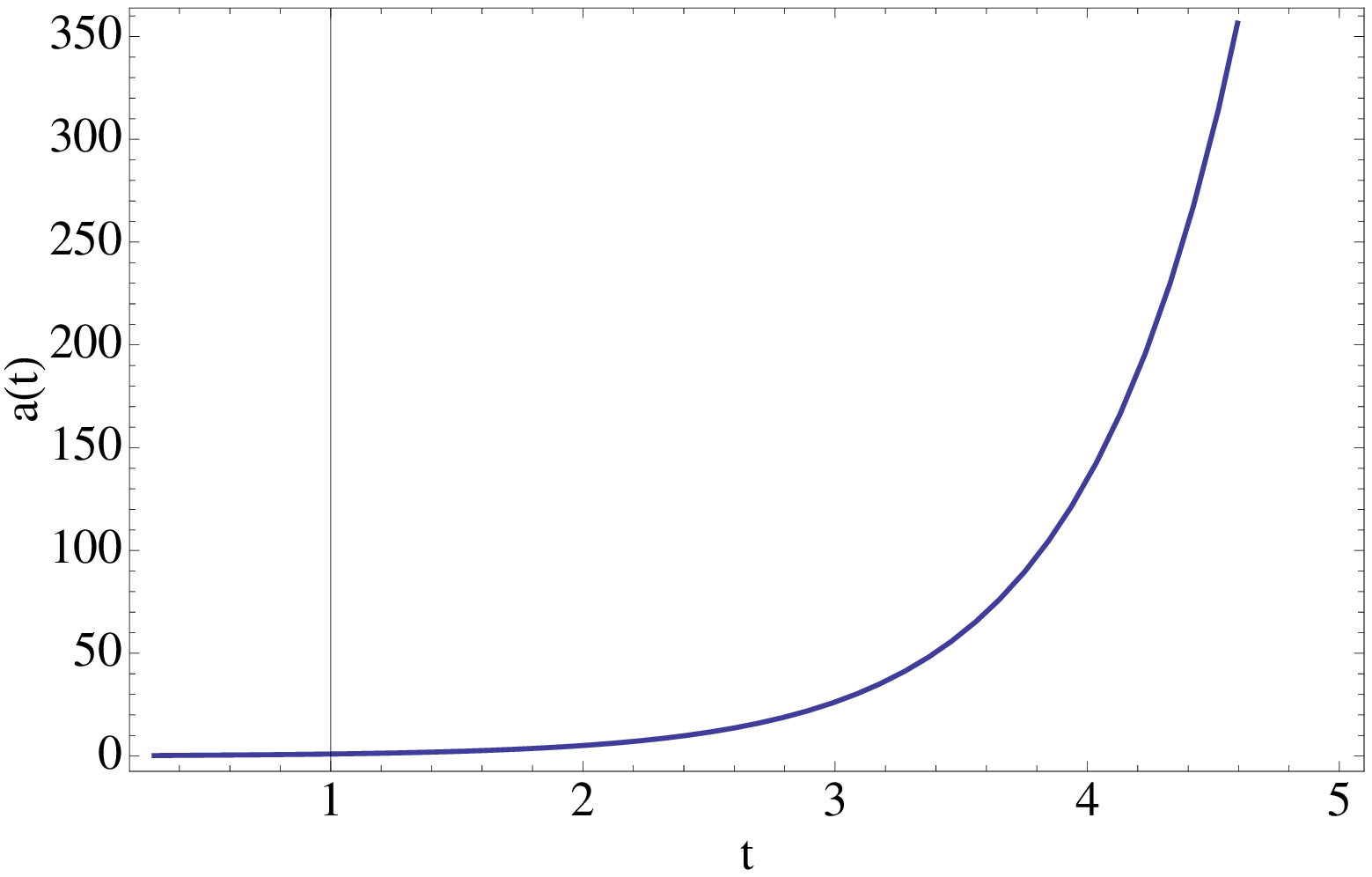}
~
\includegraphics[width=8.18cm]{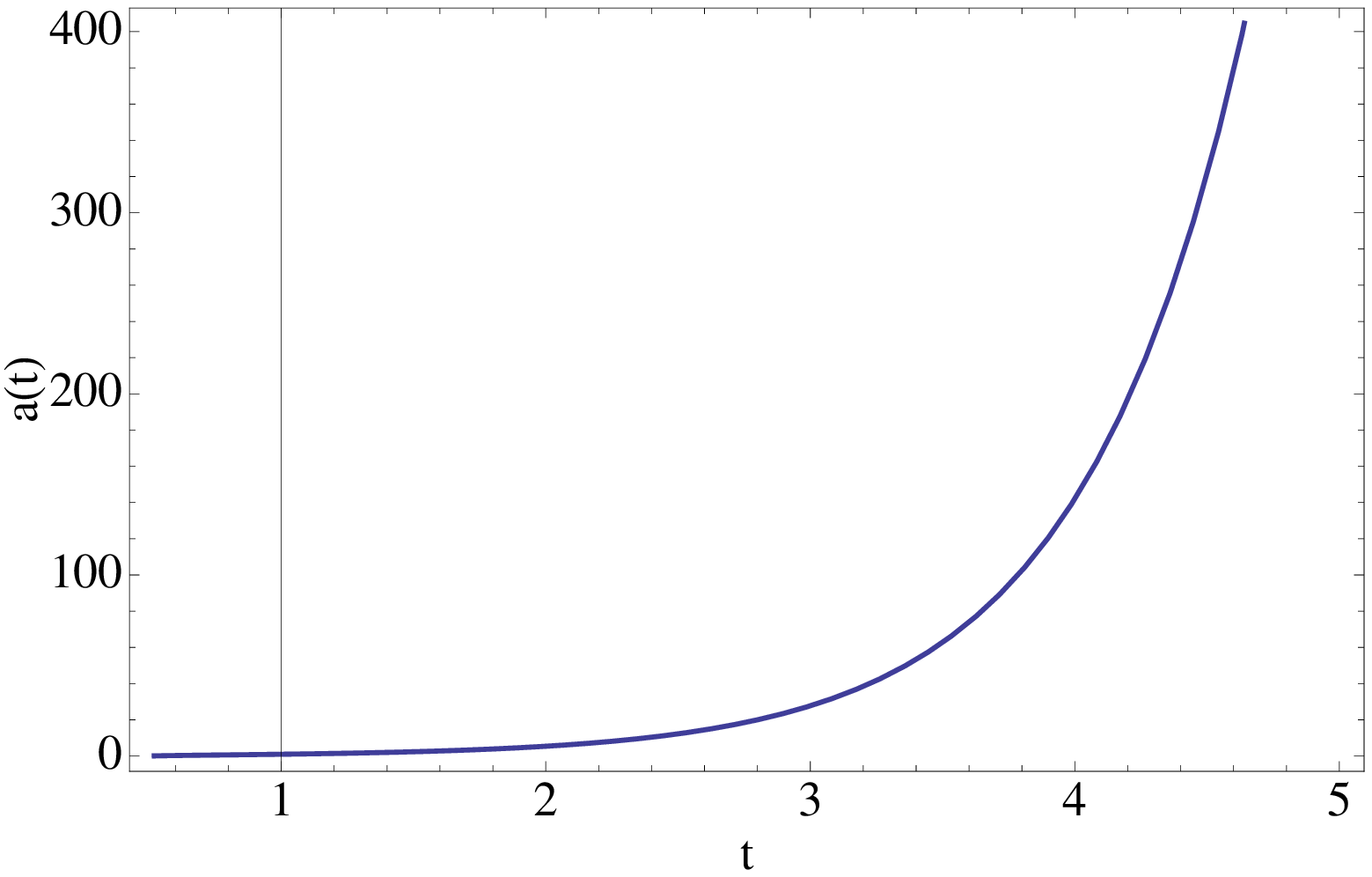}
\includegraphics[width=8cm]{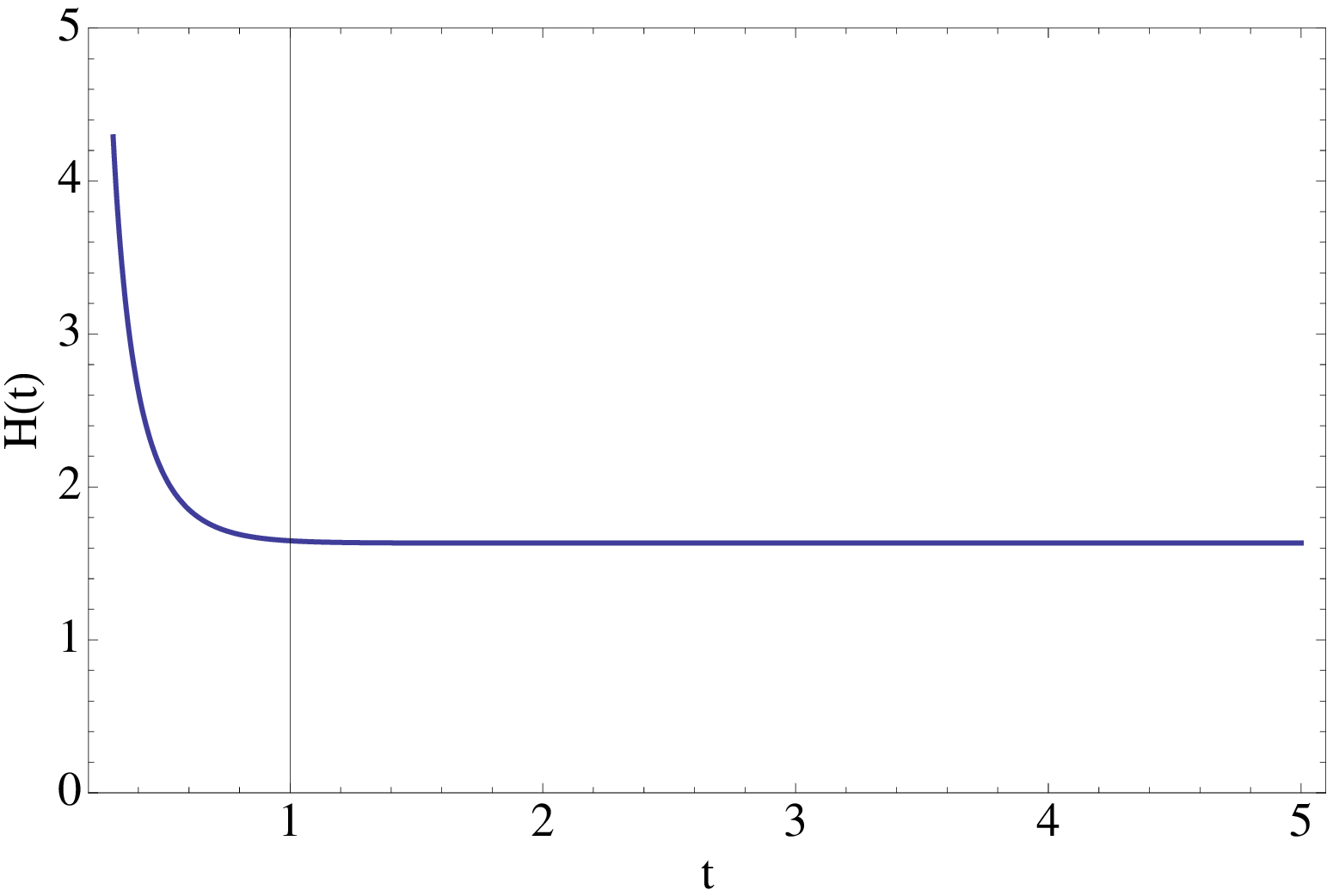}
~
\includegraphics[width=8cm]{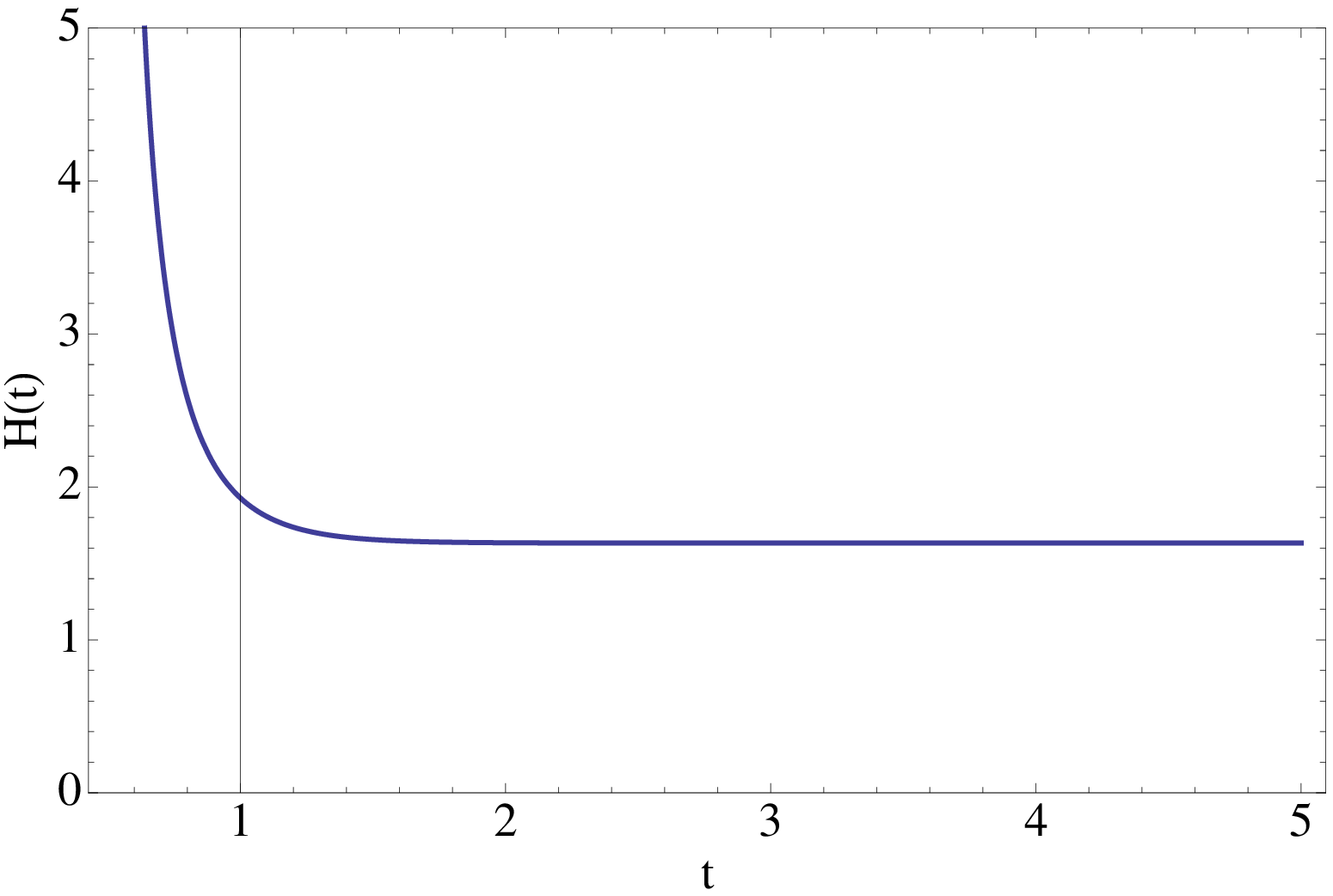}
\includegraphics[width=8cm]{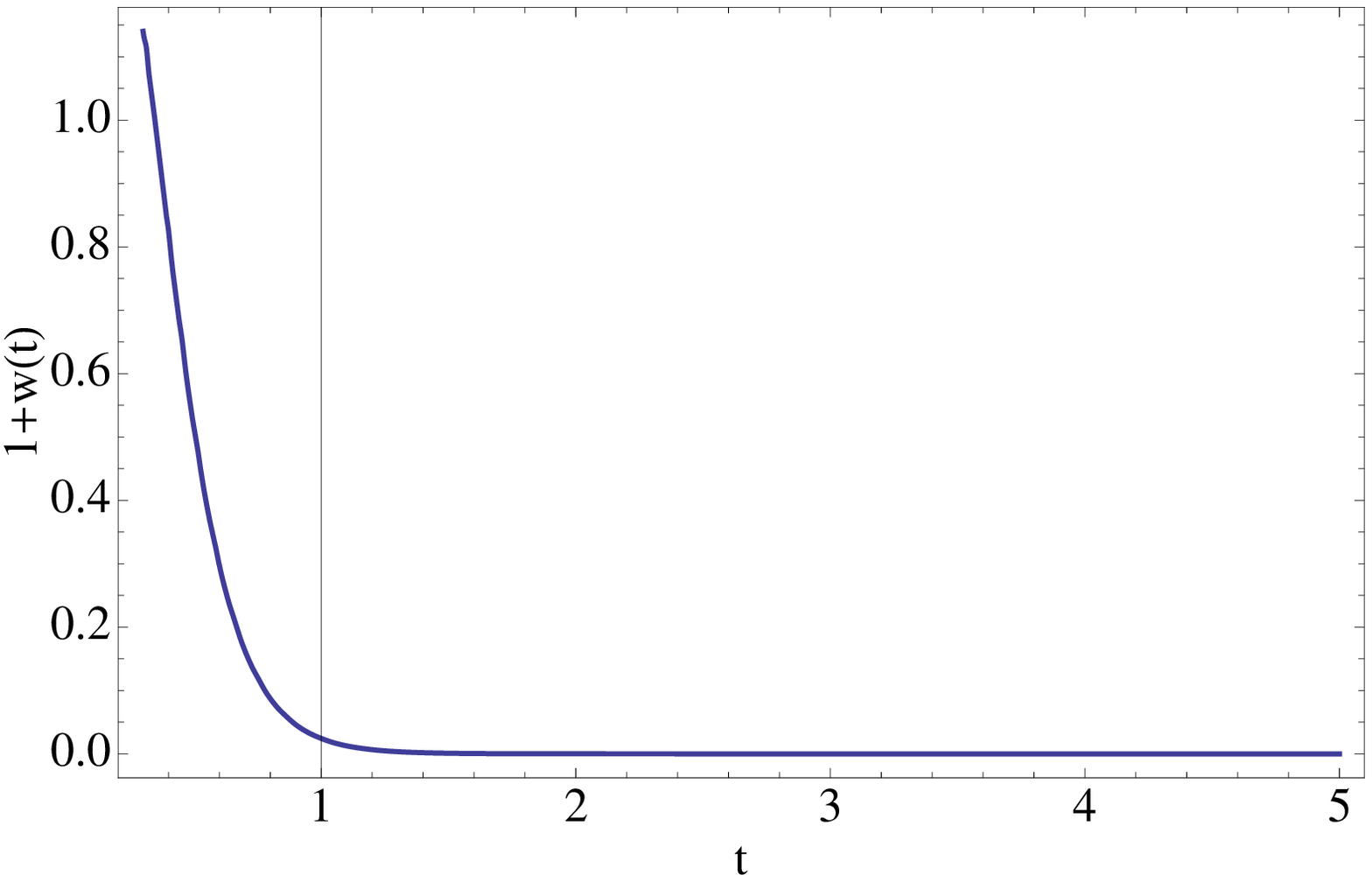}
~~~~~~
\includegraphics[width=8cm]{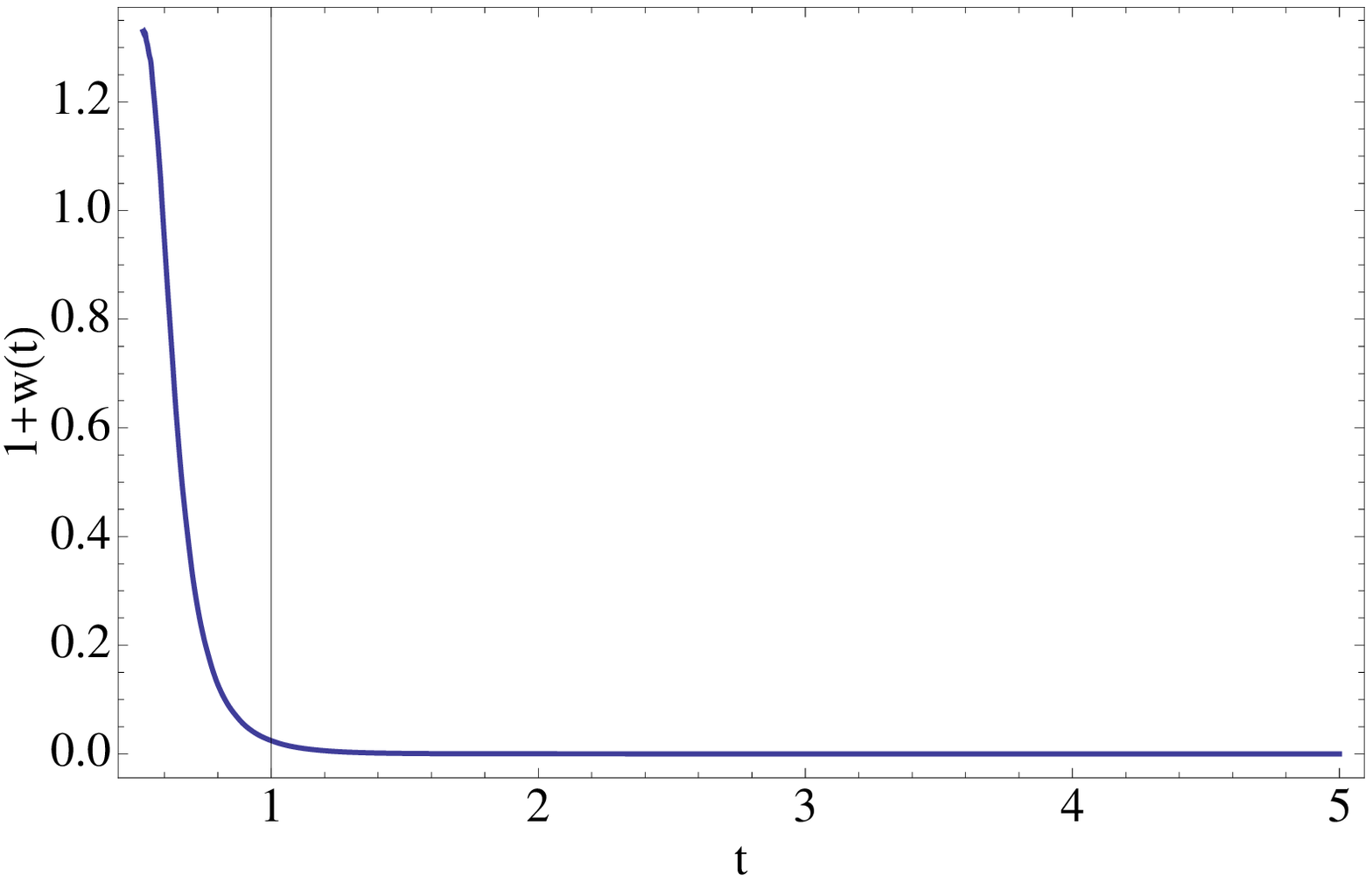}
\caption{Numerical solutions for $\zeta$, the scale factor $a$, the Hubble parameter $H$
and the equation of state $w$.}
\label{fig:sol}
\end{figure}

In the presence of a nontrivial matter density the three equations (\ref{zeta}), 
(\ref{friedman}) and (\ref{rho})  cannot be decoupled; the system can however 
be solved numerically with sufficient accuracy.
In Figure \ref{fig:sol} we show such solutions for the field $\zeta$, the scale factor $a$, 
the Hubble parameter $H$ and the equation of state parameter $w$ as a function of 
time, both in the absence of matter (left panel) and in the presence of a finite matter 
density (right panel).  While we have not enforced a physical normalization, 
these plots allow us to understand the qualitative behavior of cosmological evolution 
in the presence of the $\zeta$ field, governed by the action (\ref{fin_action}).

 
We chose $t=1$ to represent the present time\footnote{The unit time interval is given by $N G_N^{1/2}=N\tau_{\rm Plank}$, 
implying that for a realistic cosmology the number of colors $N$ in the hidden sector has to be very large. 
This is consistent with the general discussion in earlier sections, where we kept only the leading $N$ 
dependence in the orbifold theory. This also guarantees that $1/N$ effects cannot affect the running of
the double-trace coupling, in particular the fact that it becomes negative (cf. eq.~(\ref{expsol})).} and show it by the vertical line 
in all plots; as initial conditions for the scale factor we required that $a(1)=1$.
When matter density is included, we normalize it to the observed ratio 
$\rho^m(1)/\rho^\zeta(1)\simeq1/3$.

\begin{figure}
\begin{center}
\includegraphics[width=10cm]{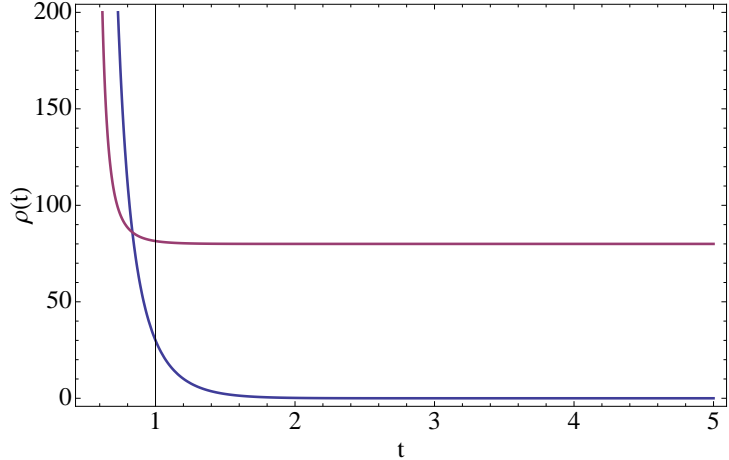}
\end{center}
\caption{The energy density of the $\zeta$ fluid (red) and matter (blue) as a function of 
time. At early times the two curves are essentially parallel, showing that the $\zeta$ 
energy density tracks the matter energy density. Despite the oscillatory behavior of $\zeta$ 
the energy densities are monotonic functions.}
\label{fig:rho}
\end{figure}

As can be seen from the figure, the field $\zeta$ oscillates at early times and eventually 
settles to a very small constant value. This behavior should indeed hold both in the 
presence and in the absence of matter, as the matter energy density diminishes at late 
times, implying that in both cases the solution should asymptote to the constant solution 
described in the previous section. 
Due to their coupling, a nontrivial matter 
density leads to an increase in early time amplitude of $\zeta$ oscillations and in 
their frequency.\footnote{While not visible in the figure, it also increases the time 
necessary for $\zeta$ to reach its asymptotic value.}
At late times, the behavior of the system is that of a cosmological constant, with a 
constant Hubble parameter, exponentially growing scale factor and equation of state
$w=-1$. At early times the equation of state of the $\zeta$ fluid varies 
with time and $w$ can reach positive values.  
In the presence of matter, the energy density in the $\zeta$ fluid partially tracks 
the matter energy density.  This is a direct consequence of the fact that the matter 
energy density sources $\zeta$, as seen in equation (\ref{zeta}). 
As we previously discussed, this is due to the quantum mechanically generated 
conformal coupling $1/6 R \zeta^2$
in (\ref{fin_action}). Fig. \ref{fig:rho} shows the energy densities in the $\zeta$ fluid 
(in red) and in matter (in blue). The position of the crossing point, shown in 
figure~\ref{fig:rho} at $a(t_c) \sim (\rho^m(1)/\rho^\zeta(1))^{1/3}$, is determined by 
the absolute values of the matter and $\zeta$ energy density.

\section{Conclusions}

In this paper we discussed a class of models that may appear if the  Standard Model 
is embedded in a more fundamental theory through a quiver gauge theory. We assumed
that, by some mechanism, these models interact only gravitationally with regular matter 
and thus have only cosmological consequences. 
The flat space properties of these models were previously considered: due to the special 
properties of their renormalization group flow, their dynamics is mainly governed by a
certain double-trace operator. Making use of this observation we constructed a simpler 
effective matrix bosonic model that captures the essential features of the original.
Its structure is such that it is completely equivalent to the quantum theory of a single 
effective scalar field. We computed the potential of this field which is 1-loop exact. Departing 
from the standard treatment of such models we also evaluated the quantum mechanically 
generated kinetic energy of this effective field and showed that it exhibits a conformal 
coupling with gravity.

The example we discussed in detail suggests that, from a cosmological standpoint, such 
almost-conformally invariant "hidden sectors" mimic the expected properties of dark 
energy at late times, while partially tracking the matter energy density at early times.
Asymptotically in the far future we found that the effective field becomes constant and 
thus its consequences are similar to those of a pure cosmological constant which is
exponentially smaller than the renormalization scale of the theory. The exponential 
suppression is governed by the coupling constant of the original double-trace operator.
The time evolution of the effective (or dark energy) field is in general nontrivial and 
sources an interesting behavior for the scale factor, Hubble parameter and equation of 
state. The value of the latter at present times is sufficiently close to $w=-1$. 
The conformal coupling plays an important role here, avoiding the usual 
extreme limits on the quintessence boson mass. In fact, the mass of the dark energy field 
is comparable to that of the far-future cosmological constant scale.

While the model we discussed in detail predicts an exponentially small far-future 
cosmological constant, we cannot, of course, claim to have solved the cosmological 
constant problem. Indeed, the contribution to the cosmological constant of the 
zero-point energy of the Standard Model fields and perhaps other hidden sectors
remains potentially uncanceled.\footnote{It is perhaps important to recall that
various mechanisms for the cancellation of  the standard model contribution to the cosmological 
constant in supersymmetry-breaking settings have been discussed in the literature -- 
see {\it e.g.} \cite{CCmess} for a partial list of references.} 
Our discussion should be thought of pointing out 
an yet unexplored source of dark energy which by itself does not introduce any 
further problems at the quantum level. 

Since the hidden sector interacts only gravitationally with regular matter, it needs not
be in thermal equilibrium with it. It is therefore possible to assume that its temperature 
vanishes, as we have done.  Perturbative inclusion of a small temperature is not difficult: 
to leading order it corrects the effective potential though a temperature-dependent mass 
term for the dark energy field $\zeta$. While the structure of the special solution discussed 
in \S4 is unchanged, the effective cosmological constant receives a temperature dependent 
positive shift. It should be possible, though perhaps not straightforward, to  analyze the 
effects of non-zero temperature beyond leading order. 

Due to the special properties of the model, a finite density of hidden quanta contributes
differently than regular matter to Einstein's equations: it is similar to a negative curvature
term without actually changing the curvature of the spatial slices. It should be interesting 
to explore the consequences of such a condensate. 

While we formulated our analysis for the special case of quiver gauge theories which 
are conformal in the planar limit, the mechanism proposed here should hold in more 
general theories in the presence of double-trace operators. While we expect the general 
features to remain the same, the technical details will likely be different and model-dependent. 
In general however, due to absence of conformal invariance,  there is 
no reason to introduce a conformal coupling for the scalar fields. It would be interesting 
to explore such more general models.

\

{\bf Acknowledgments}\\
We would like to thank  S.~Dodelson for useful comments.
This work was supported in part by the US National Science Foundation under 
PHY-0855529 and PHY-0855356, the US Department of 
Energy under contracts DE-FG02-201390ER40577 (OJI) and the 
A.P. Sloan Foundation.

\end{document}